\documentclass[11pt]{revtex4-2}
   \usepackage{relsize}
   \usepackage{braket}
   \usepackage{hyperref}
   \usepackage{amsmath,amsfonts,amssymb}
   \hypersetup{dvips,dvipdfm,colorlinks=true,urlcolor=magenta,filecolor=magenta,linktoc=page,citecolor=red,linkcolor=blue,bookmarks=true}
   \usepackage{graphicx,epstopdf}
\usepackage{bm}
\usepackage{hyperref}
\usepackage{amsmath,amsfonts,amssymb}

\hypersetup{dvips,dvipdfm,colorlinks=true,urlcolor=magenta,filecolor=magenta,linktoc=page,citecolor=red,linkcolor=blue,bookmarks=true}
\usepackage{graphicx,epstopdf}
\usepackage{array}

\newcounter{defcounter}
\newcolumntype{L}[1]{>{\raggedright\let\newline\\\arraybackslash\hspace{0pt}}m{#1}}
\newcolumntype{C}[1]{>{\centering\let\newline\\\arraybackslash\hspace{0pt}}m{#1}}
\newcolumntype{R}[1]{>{\raggedleft\let\newline\\\arraybackslash\hspace{0pt}}m{#1}}

\begin{document}
	
	\title{A complete, continuous and non-singular expansion of the Universe under stimulated  creation -annihilation process of the real scalar Bosons: An introduction to the existence of the  Anti Universe and Parallel Universe.   \\
		}
	\author{Subhayan Maity\footnote {maitysubhayan@gmail.com}}
	\affiliation{Department of Mathematics, Jadavpur University, Kolkata-700032, West Bengal, India.}
	
	\author{Manojit Das\footnote {dasmanojit430@gmail.com}}
\affiliation{Department of Physics, Raiganj University, Raiganj-733134, West Bengal, India.}

	%%%%%%%%%%%%%%%%%%%%%%%%%%%%%%%%%%%%%%%%%%%%%%%%%%%%%%%%%%%%%%%%%%%%%%%%%%%%%%%%%%%%%%%%%%%%%%%%%%%%%%%%%%%%%%%%%%%%%%
\begin{abstract}
	
	The general theory of relativity is the most popular theory to describe the dynamics of a system (especially the Universe) under gravity. In this framework, the solution of the Einstein field equation under curved space-time yields the cosmic evolution equation. Besides the evolutionary dynamics of the Universe may also be obtained from the  other aspects like thermodynamics, classical Lagrangian dynamics, symmetry analysis(Noether, Lie ) etc.
	
	This paper presents a new approach to understanding the evolution of the Universe by quantizing the cosmic fluid under gravity. While the general theory of relativity is commonly used to describe the dynamics of the Universe, this paper explores some other aspects of cosmic evolution from the particle creation-annihilation mechanism of the cosmic fluid. The model suggests that the Universe and Anti-Universe can coexist, and that there may be a parallel system (CPT-invariant) of the  Universe and Anti-Universe, all of which are created through the adiabatic particle creation-annihilation mechanism of a modified real scalar field acting as the cosmic fluid. This work provides a different approach to obtaining the cosmic evolution equation from the quantum field theory. Also the consequence of the quantization of the cosmic fluid addresses the non-singular origin of the Universe and its continuous-complete evolution.

\end{abstract}

\maketitle

%%%%%%%%%%%%%%%%%%%%%%%%%%%%%%%%%%%%%%%%%%%%%%%%%%%%%%%%%%%%%%

%%%%%%%%%%%%%%%%%%%%%%%%%%%%%%%%%%%%%%%%%%%%%%%%%%%%%%%%%%%%%%%%%%%%%%%%%%%%%%%%%%%%%%%%%%%%%%%%%%%%%%%%%%%%%%%

%\myclassification{04.70.Dy $-$ 04.60.Kz $-$ Black hole physics~~;\\$~~~~~~~~~~~~~$ 05.70.-a $-$ Thermodynamics~~;\\$~~~~~~~~~~~~~$  95.30.Tg $-$ Thermodynamic processes, equation of state~~;\\$~~~~~~~~~~~~~$  95.30.Sf $-$ Relativity and gravitation }

%\tableofcontents
%\newpage

%%%%%%%%%%%%%%%%%%%%%%%%%%%%%%%%%%%%%%%%%%%%%%%%%%%%%%%%%%%%%%%%%%%%%%
\section{Introduction}
	\begin{figure}[h]
	
	\centering
	\includegraphics*[width=1\linewidth]{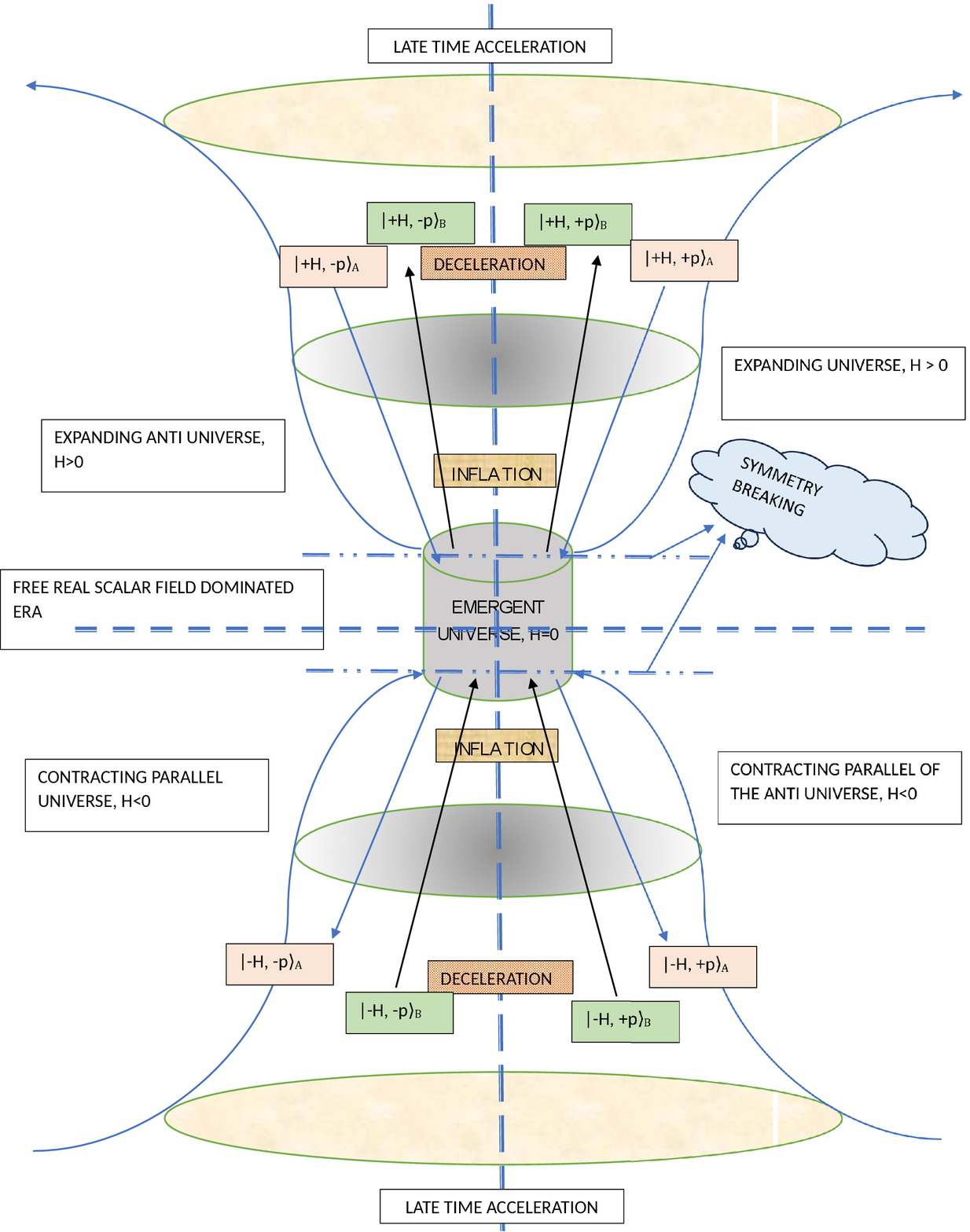}\\

	\begin{center}
		
		\caption{Particle Creation-annihilation process in the Universe + Anti Universe and Parallel Universe + Parallel of the Anti Universe.  } 
	\end{center}
	\label{fig1}
	
\end{figure}
%%%%%%%%%%%%%%%%%%%%%%%%%%%%%%%%%%%%%%%%%%%%%%%%%%%%%%%%%%%%%%%%%%%%%%%
Most of the recent researches on cosmology deal with the theoretical estimation of the  present late time acceleration as per the  observational data. People attempt to  explain the observational outcomes regarding the cosmic evolution in the light of   the general theory of relativity with some modifications ($F(R)$ model, $\Lambda$-CDM Dark fluid hypothesis, etc). But theoretically, these models suffer from several issues namely cosmological constant  problem  \cite{Weinberg:1988cp,Padmanabhan:2002ji} and coincidence problem\cite{Steinhardt:2003st}.  
\par Besides the complete and continuous cosmic evolutionary scenario \cite{Chakraborty:2014oya} has been established from the non-equilibrium thermodynamic behaviour of the Universe. These works successfully provide the best fitted cosmic evolution pattern from the phenomenological estimation \cite{Maity:2021mlx,Maity:2022gby,Maity:2022lbq,Maity:2022noc} of the thermodynamic parameters(particle creation rate, diffusion parameter , etc).  Maity in collaboration with Chakraborty, has exhibited such complete cosmic model\cite{Maity:2021mlx}  : Emergent era $\rightarrowtail$ Inflation $\rightarrowtail$ Decelerating expansion $\rightarrowtail$ Late time acceleration , from the diffusion mechanism of the cosmic fluid.  Particle creation-annihilation process is the important aspect in the cosmic non-equilibrium thermodynamics.  From the point of view of the quantum field theory, the particle creation-annihilation process \cite{Parker:1972kp} implies the canonical quantization of the cosmic fluid Hamiltonian. L.Parker in several works has shown the quantization of the Hamiltonian of the real scalar field under FLRW Universe. But the impact of this process on the cosmic evolution pattern is yet to be addressed.

\par In this work, we report the quantization process of the cosmic fluid which is compatible with the complete and continuous cosmic evolution of the universe.
Besides we have explored some interesting outcomes and insights of the adiabatic particle creation - annihilation process with some modifications in standard Lagrangian of the real scalar Bosons. 
we have demonstrated the simultaneous expansion of the Universe and a corresponding Anti Universe from the adiabatic "stimulated particle creation - annihilation" mechanism of a real scalar field fluid system. Also the parallel contracting system of this expanding (Universe+Anti-Universe) system is reported in this model.

%%%%%%%%%%%%%%%%%%%%%%%%%%%%%%%%%%%%%%%%%%%%%%%%%%%%%%%%%%%%%%%%%%%%%
\section{Quantization of the cosmic fluid in FLRW space-time.}
%%%%%%%%%%%%%%%%%%%%%%%%%%%%%%%%%%%%%%%%%%%%%%%%%%%%%%%%%%%%%%%%%%%%% 
In this work, we choose the flat FLRW space-time as the underlying geometry of the Universe. The well known form of the FLRW metric is as,
\begin{equation}
	ds^2=dt^2-a^2(t)(dx^2+dy^2+dz^2) .  \label{1}
\end{equation}
Here the natural unit system is preferred by taking the Universal constants as unity, $G=c=1$.
$a(t)$ is the time dependent scale factor of the Universe. Following the Einstein's framework of General Relativity (GR), one has the friedmann equations
\begin{equation}
	3H^2= 8 \pi \rho, 2\dot{H}=-8\pi( P+\rho) . \label{2}
\end{equation}
$H=\frac{\dot{a}}{a}$ is the Hubble parameter. $P$ and $\rho$ are the thermodynamic pressure and density of the cosmic fluid (single fluid model of the Universe with isotropic ideal gas like fluid $T^{\mu \nu}=\mbox{Dia}(\rho,P,P,P)$) respectively. These friedmann equations effectively yield the evolution equation of the Universe as \begin{equation}
	\frac{2\dot{H}}{3H^2}+\frac{P+\rho}{\rho}=0. \label{3}
\end{equation}
These are also compatible with the first law of thermodynamics 
\begin{equation}
	\dot{\rho}+3H (P+\rho)=0 . \label{4}
\end{equation}
Equation(\ref{3}) and (\ref{4}) respectively describe the dynamical and thermodynamic aspects of an isolated Universe. The framework of GR is purely classical and arguably incompatible with the early time evolution of the Universe(in smaller dimension). 
\par In this work, we report a cosmic evolution equation from the quantum field theoretical point of view. We choose the cosmic fluid of real scalar field particle with Lagrangian density (in flat Minkowski space-time) 
\begin{equation}
	\mathcal{L}_{Minkwoski}=\frac{1}{2}\partial_{\mu} \phi \partial ^{\mu} \phi- \frac{1}{2} m^2 \phi ^2  , \label{5}
\end{equation} 
$\phi =\phi ^*$, a real scalar field and $m$ is the mass of a cosmic fluid particle. Under any curved space-time, the Lagrangian density will be modified as 
\begin{equation}
		\mathcal{L}= \frac{1}{2} \nabla_{\mu}\phi\nabla^{\mu}\phi -\frac{1}{2} (m^2+\zeta R) \phi ^2 , \label{6}
\end{equation} 
$\nabla_{\mu}$ represents the covariant derivative. $R$ is the Ricci scalar and $\zeta$ is the coupling parameter. Here we choose $\zeta =0$ for minimal coupling. Following the Euler- Lagrangian equation (EL) one finds the Klein-Gordon (KG) equation as
\begin{equation}
		(\Box ^2 +m^2) \phi =0,  \label{7}
\end{equation}
where $\Box^2$ represents the d'alembertian operator. Under flat FLRW metric (\ref{1}), one obtains the explicit form of the KG equation as,
\begin{equation}
	\ddot{\phi}+3H \dot{\phi} - \nabla^{\prime  ^2} \phi+ m^2 \phi =0 .\label{8}
\end{equation} Here $\vec{\nabla}^{\prime}=\frac{1}{a}\vec{\nabla}$, the co-moving Laplacian operator. The first order derivative term $\dot{\phi}$ indicates the dissipation of energy from the fluid system. On the other hand the KG equation in Minkowski space-time corresponds to the free harmonic oscillator. Thus one can conclude that the effect of curvature in space-time(here FLRW) is to accomplish dissipation from the source(fluid). However in order to quantize the system, one has to assume the solution of the KG equation as the superposition of the infinite number of harmonic oscillators with different momenta ranging $-\infty < k < +\infty $. Here we choose the trial solution of the equation (\ref{8}) as,
\begin{equation}
	\phi (X)\sim\int d^3k \tilde{\phi}(K)e^{-i\int \tilde{K}dX} . \label{9}
\end{equation}
$\tilde{K}=\tilde{k}^{\mu}=(k^0,\vec{k})$, the effective four momenta in this case. Hence one may write the form of the real scalar field $\phi (X)$ as,
\begin{equation}
		\phi(X)=\frac{1}{(2 \pi)^{\frac{3}{2}}} \int\frac{1}{\sqrt{2 \omega_0}} d^3k^{\prime}\left[ \mathcal{A}(\vec{k}^{\prime},t) e^{-i\int KdX} +\mathcal{A}^* (\vec{k}^{\prime},t)e^{+i\int KdX}  \right], \label{10}
\end{equation} 
where $ \mathcal{A}(\vec{k^{\prime}},t)=e^{-\frac{3}{2}\int Hdt} \alpha (\vec{k^{\prime }})$ and  $ \mathcal{A}^*(\vec{k^{\prime}},t)=e^{-\frac{3}{2}\int Hdt} \alpha^* (\vec{k^{\prime }})$. Also $k^0=i\frac{3}{2}H \pm \omega (t)$, $\int KdX= \int \omega (t) dt - \vec{k}.\vec{x}$. \par Here we assume the slow variation of $k^0$ i.e. $
\frac{\dot{k}^0}{k^0}<<1$.    One may introduce $\vec{x}\rightarrow \vec{x}^{\prime}=a \vec{x}, \vec{k}\rightarrow \vec{k}^{\prime}=\frac{\vec{k}}{a}$. $\vec{x}^{\prime}, \vec{k}^{\prime}$ are the comoving space-coordinate and momenta respectively.Eventually $\vec{k}.\vec{x}= \vec{k}^{\prime}.\vec{x}^{\prime}$.  Here $\omega(t)= \sqrt{|\omega_0 ^2 - \frac{9}{4} H^2|},  ~\omega_0=\sqrt{k^{\prime 2}+m^2}$ . Evidently, The solution of K-G equation is the  superposition of infinite numbers of damped harmonic oscillators with momentum values ranging $-\infty <k^{\prime}< \infty$.
 \par The existence of the damping in the solution of the K-G equation justifies the dissipation of the energy of the scalar field in to the  cosmic expansion energy. Similarly one may think of the reverse process i.e. the dissipation of energy into the scalar field. In that case, the solution of K-G equation will contain the time growing harmonic oscillators.  However, in this work we have aimed to consider the interaction between the cosmic evolution process and the dynamics of the real scalar field Lagrangian in both these possible ways. Physically it will yield the evolution of the combination of the Universe and its Anti- Universe. 
 \par CPT (simultaneous operation of the Charge conjugation operator, Parity and time reversal operator ) invariance is the basic symmetry of elementary particles which preserves the Hamiltonian of the system. Under CPT operation, a system transforms into its corresponding "parallel system" which follows identical laws of physics.

 \begin{equation*}
 	\mbox{Universe}~~~  \underleftrightarrow{\mbox{CPT}}~~~ \mbox{Parallel Universe}.
 \end{equation*}
Again one has $ e^{-\frac{3}{2}\int Hdt} e^{-i\int KdX}\underleftrightarrow{\mbox{CPT}}~~~e^{-\frac{3}{2}\int Hdt}  e^{+i\int KdX} $. Thus both the Universe and the corresponding Parallel Universe can be governed by the damped harmonic oscillator scalar field.
 One has a contracting Parallel Universe of an expanding Universe and an expanding Parallel Universe (PU) of a contracting Universe.
\begin{equation*}
		\mbox{Universe}~  \xrightarrow{t\rightarrow-t, H\rightarrow H^{\prime}= -H}~ \mbox{Parallel Universe} \xrightarrow{t\rightarrow-t, H^{\prime}\rightarrow -H^{\prime}=H} \mbox{ Anti Universe}.
\end{equation*} 
As $ e^{-\frac{3}{2}\int Hdt} \alpha (\vec{k^{\prime }})e^{-i\int KdX}\underleftrightarrow{t \rightarrow-t, H\rightarrow H}~~~e^{+\frac{3}{2}\int Hdt} \alpha (\vec{k^{\prime }}) e^{+i\int KdX} $, the time growing Harmonic oscillator scalar field will correspond to the evolution of the Anti-Universe (AU).

Here we infer  
 \begin{equation} 
 	\ket{t,H}_{\mbox{System}}=\ket{t,H}_{\mbox{Universe}}+\ket{t,H}_{\mbox{ Anti Universe}},   \label{11}
 \end{equation}
where $\ket{t,H}_{\mbox{ Anti Universe}}, \ket{t,H}_{\mbox{Universe}}$, both follow identical evolution with Hubble parameter $H$. Therefore the wave function of the system represents the state of the combination of the Universe and its corresponding  Anti Universe (AU). This system (Universe-Anti Universe) has a Parallel system with Hubble parameter ($-H$) as
$\ket{t,H}_{\mbox{Parallel System}}=\ket{t,-H}_{\mbox{Universe}}+\ket{t,-H}_{\mbox{ Anti Universe}}$' 
\begin{eqnarray*}
		\ket{t,H}_{\mbox{System}}\underleftrightarrow{\mbox{CPT}}\ket{t,-H}_{\mbox{ Parallel System}}, \\
			\ket{t,H}_{\mbox{Universe}}\underleftrightarrow{\mbox{CPT}}\ket{t,-H}_{\mbox{ Parallel Universe}},\\
				\ket{t,H}_{\mbox{Anti Universe}}\underleftrightarrow{\mbox{CPT}}\ket{t,-H}_{\mbox{ Parallel of the Anti Universe}}
\end{eqnarray*}
  Hence we have assumed both the damped and time growing oscillator  real scalar field for the corresponding Lagrangian. This proposed solution will not satisfy the K-G equation (\ref{8}). One can modify the equation of motion as(by introducing an extra source term),
  \begin{equation}
  		\ddot{\phi}+3H \dot{\phi} - \nabla^{\prime  ^2} \phi+ m^2 \phi =\chi(t).  \label{12}
  \end{equation}
Hence the modified system will be a forced vibrator under an external source $\chi(t)$. $\chi(t)$  is the analog of the external force density of the forced vibration. Hence the consequent system will not be the conventional real scalar field system rather it will turn in to a modified system where the Lorentz symmetry of the Lagrangian will be spontaneously broken. The impact of this mechanism will be discussed later in this article (Section $2.3$). However the canonical quantization of the Hamiltonian of this modeled system corresponds to the stimulated  creation-annihilation mechanism of the real scalar bosons due to the source $\chi(t)$.  

\par For the desired solution, we have chosen(phenomenological) the periodic source term as,  \par 
$\chi(t)=2 \frac{1}{(2 \pi)^{\frac{3}{2}}}\left [\omega ^2+\frac{9}{2}H^2+i3\omega H \right ] \int\frac{1}{\sqrt{2 \omega_0}} d^3k^{\prime}\left[ \alpha (\vec{k^{\prime}},t) e^{-i\int KdX} +\alpha ^{*} (\vec{k^{\prime}},t)e^{+i\int KdX}  \right]e^{\frac{3}{2}\int H dt}$.
\par Hence the general solution can be written as,
\begin{equation}
	\phi(X)=\phi_{+}(X)+\phi_{-}(X), \label{13}	
\end{equation}
where $\phi_{-}(X)=\frac{1}{(2 \pi)^{\frac{3}{2}}} \int\frac{1}{\sqrt{2 \omega_0}} d^3k^{\prime}\left[ A(\vec{k^{\prime}},t) e^{-i\int KdX} +A^* (\vec{k^{\prime}},t)e^{+i\int KdX}  \right]$ and \par $\phi_{+}(X)=\frac{1}{(2 \pi)^{\frac{3}{2}}} \int\frac{1}{\sqrt{2 \omega_0}} d^3k^{\prime}\left[ B(\vec{k^{\prime}},t) e^{-i\int KdX} +B^* (\vec{k^{\prime}},t)e^{+i\int KdX}  \right]$.  \par $ A(\vec{k^{\prime}},t)=e^{-\frac{3}{2}\int Hdt} \alpha (\vec{k^{\prime}})$,  \par  
$ B(\vec{k^{\prime}},t)=e^{+\frac{3}{2}\int Hdt} \alpha (\vec{k^{\prime}})$.
\par The corresponding quantum field operator is in the form
\begin{equation}
	\hat{\phi}(X)=\hat{\phi}_{+}(X)+\hat{\phi}_{-}(X), \label{14}
\end{equation}
where $\hat{\phi}_{-}(X)=\frac{1}{(2 \pi)^{\frac{3}{2}}} \int\frac{1}{\sqrt{2 \omega_0}} d^3k^{\prime}\left[ \hat{A}(\vec{k^{\prime}},t) e^{-i\int KdX} +\hat{A}^{\dagger}  (\vec{k^{\prime}},t)e^{+i\int KdX}  \right]$ and \par $\hat{\phi}_{+}(X)=\frac{1}{(2 \pi)^{\frac{3}{2}}} \int\frac{1}{\sqrt{2 \omega_0}} d^3k^{\prime}\left[\hat{ B}(\vec{k^{\prime}},t) e^{-i\int KdX} +\hat{B}^{\dagger} (\vec{k^{\prime}},t)e^{+i\int KdX}  \right]$.
\par The corresponding Hamiltonian in this case is given by
\begin{eqnarray}
	h=\frac{1}{2}\int d^3 x^{\prime} \left[{\dot{
			{
				\phi}}}^2+|\vec{\nabla}^{\prime}\phi|^2\ + m^2 \phi ^2 \right].  \label{15}
\end{eqnarray}
In this case, the normal ordered Hamiltonian operator is found in the form, 

\begin{equation} 
	:\hat{h}(X):=\hat{h}_0+3H \sqrt{\frac{9}{4}H^2+\omega ^2} \left[e^{{i\tan ^{-1}\frac{2\omega }{3H}}}\hat{h}_1 + e^{{-i\tan ^{-1}\frac{2\omega }{3H}}}\hat{h}_2 \right]+\omega \sqrt{4\omega ^2 +9 H^2}\left[ e^{{i\tan ^{-1}\frac{3H }{2\omega}}}\hat{h}_3  +e^{{-i\tan ^{-1}\frac{3H }{2\omega}}} \hat{h}_4 \right] . \label{16}
\end{equation}  
Here 
\begin{eqnarray}
	\hat{h}_0& =&\int d^3 k^{\prime} ~ \omega_0 \left[\hat{A}^{\dagger}(\vec{k^{\prime}},t)\hat{A}(\vec{k^{\prime}},t)+\hat{B}^{\dagger}(\vec{k^{\prime}},t)\hat{B}(\vec{k^{\prime}},t)\right] \label{17}\\
	\hat{h}_1& =&\int d^3 k^{\prime} ~ \frac{1}{2 \omega_0} \left[\hat{A}(\vec{k^{\prime}},t)\hat{A}(-\vec{k^{\prime}},t) e^{-2i\int \omega dt}+\hat{B}^{\dagger}(\vec{k^{\prime}},t)\hat{B}^{\dagger}(-\vec{k^{\prime }},t)e^{2i \int \omega dt} \right]   \label{18} \\
	\hat{h}_2& =&\int d^3 k^{\prime} ~ \frac{1}{2\omega_0} \left[\hat{B}(\vec{k^{\prime}},t)\hat{B}(-\vec{k^{\prime}},t) e^{-2i\int \omega dt}+\hat{A}^{\dagger}(\vec{k^{\prime}},t)\hat{A}^{\dagger}(-\vec{k^{\prime}},t)e^{2i \int \omega dt} \right]  \label{19} \\
		\hat{h}_3& =&\int d^3 k^{\prime} ~\frac{1}{2 \omega_0}  \left[\hat{B}^{\dagger}(\vec{k^{\prime}},t)\hat{A}(\vec{k^{\prime}},t)\right] \label{20} \\
		\hat{h}_4& =&\int d^3 k^{\prime} ~ \frac{1}{2\omega_0} \left[\hat{A}^{\dagger}(\vec{k^{\prime}},t)\hat{B}(\vec{k^{\prime}},t)\right] .\label{21} 
\end{eqnarray}

Evidently, $\hat{h}_0$ is readily in canonically quantized form with two sets of creation and annihilation operators. $\hat{A}^{\dagger}(\vec{k^{\prime}},t), \hat{A}(\vec{k^{\prime }},t)$ are the ladder operators for type - A particles (say) and $\hat{B}^{\dagger}(\vec{k^{\prime}},t), \hat{B}(\vec{k^{\prime}},t)$ for type - B particles.  A-type particle dissipates energy to the Universe and B-type particle acquires energy from the Universe.
The terms containing $\hat{h}_1,\hat{h}_2,\hat{h}_3$ and $\hat{h}_4$ are the  non-trivial terms.

Now we introduce a Bogolubov transformation as 
\begin{equation}
	\hat{\mathcal{F}}_A=\alpha _1\hat{A}(\vec{k^{\prime},t}) +\beta_1\hat{A}^{\dagger}(-\vec{k^{\prime},t})   \label{22}
\end{equation}  and
\begin{equation}
	\hat{\mathcal{F}}_B=\alpha _2\hat{B}(\vec{k^{\prime},t}) +\beta_2\hat{B}^{\dagger}(-\vec{k^{\prime},t}) . \label{23}
\end{equation}
$\alpha_1,\alpha_2,\beta_1$ and $\beta_2$ are the bogolubov coefficients. 
\par Let $|\alpha_i|^2-|\beta_i|^2=1,(i=1,2)$ for satisfying the canonical transformation \par  $[\hat{A}^{\dagger}(\vec{k^{\prime},t}),\hat{A}(\vec{k,t})]=\delta _{k,k^{\prime}}=[\hat{\mathcal{F}}^{\dagger}_A(\vec{k^{\prime}}),\hat{\mathcal{F}}_A(\vec{k})]$ and $[\hat{B}^{\dagger}(\vec{k^{\prime},t}),\hat{B}(\vec{k^{\prime},t})]=\delta_{k,k^{\prime}}=[\hat{\mathcal{F}}^{\dagger}_B(\vec{k^{\prime}}),\hat{\mathcal{F}}_B(\vec{k})]$. \par  Notably for Bosons, the commutation of the Creation and Annihilation operator is $1$.
\par $\epsilon \alpha_1 ^{*}\beta_1 =\epsilon\alpha_2 \beta_2^{*}=\frac{3H}{2 \omega_0^2}\left(\frac{3}{2}H-i\omega\right) ,~ \epsilon  \beta_1 ^{*}\alpha_1 =\epsilon\beta_2 \alpha_2^{*}=\frac{3H}{2 \omega_0^2}\left(\frac{3}{2}H+i\omega\right)$. $\epsilon$ is a pure constant. Then the form of the field operator can be written in the form
\begin{eqnarray}
\hat{\phi}_{-}(X)=\frac{1}{(2 \pi)^{\frac{3}{2}}} \int\frac{1}{\sqrt{2 \omega_0}} d^3k^{\prime}\left[ 	\hat{\mathcal{F}}_A  e^{-i\int KdX} +\hat{\mathcal{F}}^{\dagger}_A  (\vec{k^{\prime}},t)e^{+i\int KdX}  \right]	\label{24} \\
\hat{\phi}_{+}(X)=\frac{1}{(2 \pi)^{\frac{3}{2}}} \int\frac{1}{\sqrt{2 \omega_0}} d^3k^{\prime}\left[ 	\hat{\mathcal{F}}_B e^{-i\int KdX} +\hat{\mathcal{F}}^{\dagger} _B (\vec{k^{\prime}},t)e^{+i\int KdX}  \right] \label{25} \\
\mbox{with}~	\hat{\phi}(X)=\hat{\phi}_{+}(X)+\hat{\phi}_{-}(X)  \label{26}
\end{eqnarray}
Hence the normal ordered Hamiltonian can be rewritten in the form

\begin{align*}
		:\hat{h}: =\int d^3 k^{\prime} ~\epsilon \omega_0 \left[\hat{\mathcal{F}}_A ^{\dagger}(\vec{k^{\prime}},t)\hat{\mathcal{F}}_A (\vec{k^{\prime}},t)+\hat{\mathcal{F}}_B ^{\dagger}(\vec{k^{\prime}},t)\hat{\mathcal{F}}_B (\vec{k^{\prime}},t)\right] \\ +\frac{\omega}{2 \omega_0} \sqrt{4 \omega ^2 +9H^2}\int d^3 k^{\prime}  \left[\hat{B} ^{\dagger}(\vec{k^{\prime}},t)\hat{A} (\vec{k^{\prime}},t)e^{{-i\tan ^{-1}\frac{3H }{2\omega}}}+\hat{A} ^{\dagger}(\vec{k^{\prime}},t)\hat{B}_B (\vec{k^{\prime}},t) e^{{i\tan ^{-1}\frac{3H }{2\omega}}}\right],
\end{align*} with $\epsilon (|\alpha_i|^2+|\beta_i|^2)=1$.
\subsection{Conditions for canonical quantization
}
One may find several conditions to quantize the Hamiltonian. 
\par $(i)$ $H=0, a\neq 0$.
\par In an emergent model of the Universe\cite{Mukherjee:2005zt,Mukherjee:2006ds,Beesham:2009zw,Paul:2020bje,Zhang:2013ykz,Paul:2015eja,Debnath:2017xcu,Paul:2018ppy,Debnath:2020bno,Debnath:2021ncz,Paul:2022dsb,Paul:2021lvb,Paul:2010jb,Paul:2011nw,Ghose:2011fk,Labrana:2013oca,Paul:2019oxo}, one has at the past infinity ($t\rightarrow -\infty$); $H=0,a\rightarrow a_E$(a non-zero constant) and  $\omega \rightarrow \omega _0$. At this epoch of time, the normal ordered Hamiltonian of the system approaches to the form
\begin{equation}
	:\hat{h}_E:=\int d^3 k^{\prime} ~ \omega_0\left (a_E^3+a_E^{-3}+1\right ) \left[\hat{\alpha}^{\dagger}(\vec{k^{\prime}},t)\hat{\alpha}(\vec{k^{\prime}},t)\right] .  \label{27}
\end{equation}
Here $a_E$ is the constant scale factor at the epoch of origin of an emergent Universe. $\omega_0\neq 0$ must be satisfied because it will correspond to the non-zero vacuum energy of the system. The emergent scenario is relevant in this model to avoid the ultraviolet divergence of the Hamiltonian at past infinity. However this model accepts the very small value of $a_E$ i.e. $a_E \rightarrow0$ but $a_E \neq 0$, the vacuum energy at the origin of the Universe will be very high. Notably, one may write
\begin{equation}
		:\hat{h}_E:=\left (a_E^3+a_E^{-3}+1\right ):\hat{h}:_{\mbox{Minkowski}}, \label{28}
\end{equation}
where $:\hat{h}:_{\mbox{Minkowski}}$ is the normal ordered Hamiltonian of the real scalar field system in Minkowski space-time (flat).
\par $(ii)$ $H\neq 0, \omega =0$
\par In this condition, the Hamiltonian of the system can be quantized by the Bogolubov transformed creation- annihilation operators ($\hat{\mathcal{F}}^{\dagger}_A,\hat{\mathcal{F}}_A$) and ($\hat{\mathcal{F}}^{\dagger}_B,\hat{\mathcal{F}}_B$). The form of the Hamiltonian operator will be 
\begin{equation}
	:\hat{h}: =\int d^3 k^{\prime} ~ \epsilon\omega_0 \left[\hat{\mathcal{F}}_A ^{\dagger}(\vec{k^{\prime}},t)\hat{\mathcal{F}}_A (\vec{k^{\prime}},t)+\hat{\mathcal{F}}_B ^{\dagger}(\vec{k^{\prime}},t)\hat{\mathcal{F}}_B (\vec{k^{\prime}},t)\right] . \label{29}	
\end{equation}
But the only two values of $\omega_0=\pm \frac{3}{2}H$ with same momentum values $k^{\prime}=\pm\sqrt{\frac{9}{4}H^2-m^2}=\pm p$ (say) are valid in this canonical quantization process.  One also has  $\epsilon \alpha_1 ^{*}\beta_1 =\epsilon\alpha_2 \beta_2^{*}=1 ,~ \epsilon  \beta_1 ^{*}\alpha_1 =\epsilon\beta_2 \alpha_2^{*}=1$ in the limit $\omega =0$. Thus the numerical value of $\epsilon$ is found to be $\sqrt{5}$.
\subsection{CPT invariance : concept of the Parallel Universe } The positive and negative energy eigen values with same magnitude is associated with the existence of the system and its parallel system (CPT invariant).  
The energy of a particle is $+\frac{3\sqrt{5}}{2}H$   in an expanding Universe($H>0$) with momentum $k^{\prime}=+\sqrt{\frac{9}{4}H^2-m^2}=+p$ and simultaneously in the expanding Anti Universe the energy of a corresponding particle is $+\frac{3\sqrt{5}}{2}H$ with momentum value $k^{\prime}=-\sqrt{\frac{9}{4}H^2-m^2}=-p$. Similarly in the contracting Parallel Universe($H<0$), the energy of a particle is   
$-\frac{3\sqrt{5}}{2}H$ with momentum $k^{\prime}=-\sqrt{\frac{9}{4}H^2-m^2}=-p$ while in the parallel of the Anti Universe, the energy of each particle is $-\frac{3\sqrt{5}}{2}H$ with momentum $k^{\prime}=+\sqrt{\frac{9}{4}H^2-m^2}=+p$. For contracting ( Universe + Anti - Universe ) system and its parallel system the sign of the energy eigen value will be reversed with no change in the momentum. Notably the energy eigen value is positive in both the system and the parallel system. 
\par However there are two types of particles with Ladder operators ($\hat{\mathcal{F}}^{\dagger}_A,\hat{\mathcal{F}}_A$) for A-type particles and ($\hat{\mathcal{F}}^{\dagger}_B,\hat{\mathcal{F}}_B$) for B-type particles (say).

\par Single particle state of each type of particle has only two allowed energy eigen values $\pm \frac{3\sqrt{5}}{2}H$. Hence effectively one has $4$ creation operators as well as $4$ annihilation operators. $\hat{\mathcal{F}}^{\dagger}_{A\pm}$ represent the creation operators of A-type particle with energy $\pm \frac{3\sqrt{5}}{2}H$ respectively and $\hat{\mathcal{F}}^{\dagger}_{B\pm}$ represent the creation operators of B-type particle with energy $\pm \frac{3\sqrt{5}}{2}H$ respectively. Let $\ket{0}$ is the vacuum or zero particle state of the system.
\begin{eqnarray}
\hat{\mathcal{F}}^{\dagger}_{A\pm} \ket{0}=\ket{\pm H,\pm p}_A \label{30}	\\
\hat{\mathcal{F}}^{\dagger}_{B\pm} \ket{0}=\ket{\pm H,\pm p}_B \label{31}\\
\hat{\mathcal{F}}_{A\pm} \ket{0}=\hat{\mathcal{F}}_{B\pm} \ket{0}=0 \label{32}\\
:\hat{h}: \ket{\pm H,\pm p}_A =\pm \frac{3\sqrt{5}}{2}H  \ket{\pm H,\pm p}_A \label{33} \\
:\hat{h}: \ket{\pm H,\pm p}_B =\pm \frac{3\sqrt{5}}{2}H  \ket{\pm H,\pm p}_B,   \label{43}
\end{eqnarray}
where $\ket{\pm H,\pm p}_A,\ket{\pm H,\pm p}_B $ are the single particle state of A and B type particles in two systems (real and parallel ) with positive momenta in the Universe and parallel of Anti Universe and negative momenta in the Anti Universe and the Parallel Universe  respectively.
\begin{equation*}
	\ket{\pm H,\pm p}_A \xrightarrow[\mbox{Parallel Universe $\rightarrowtail$ Parallel of Anti Universe}]{\mbox{Universe $\rightarrowtail$ Anti Universe}} \ket{\pm H, \mp p}_B.  
\end{equation*}
\begin{equation*}
	\ket{\pm H,\pm p}_B \xrightarrow[\mbox{Parallel Universe $\rightarrowtail$ Parallel of Anti Universe}]{\mbox{Universe $\rightarrowtail$ Anti Universe}} \ket{\pm H, \mp p}_A.  
\end{equation*}
\begin{equation*}
	\ket{\pm H,\pm p}_A \xrightarrow[\mbox{Anti Universe $\rightarrowtail$ Parallel of the Anti Universe}]{\mbox{ Universe $\rightarrowtail$ Parallel  Universe}} \ket{\mp H, \mp p}_A.  
\end{equation*}
\begin{equation*}
	\ket{\pm H,\pm p}_B \xrightarrow[\mbox{Anti Universe $\rightarrowtail$ Parallel of the Anti Universe}]{\mbox{ Universe $\rightarrowtail$ Parallel  Universe}} \ket{\mp H, \mp p}_B.  
\end{equation*}
Hence one can report that
\begin{eqnarray}
	:\hat{h}:_{\mbox{Universe}}=  \pm \frac{3\sqrt{5}}{2} H \left[\hat{\mathcal{F}}_{A+} ^{\dagger}\hat{\mathcal{F}}_{A+} +\hat{\mathcal{F}}_{B+} ^{\dagger}\hat{\mathcal{F}}_{B+} \right] \label{35} \\
	:\hat{h}:_{\mbox{Parallel Universe}}=\mp\frac{3\sqrt{5}}{2} H\left[\hat{\mathcal{F}}_{B-} ^{\dagger}\hat{\mathcal{F}}_{B-}+\hat{\mathcal{F}}_{A-} ^{\dagger}\hat{\mathcal{F}}_{A-} \right]  \label{36} \\
	:\hat{h}:_{\mbox{ Anti Universe}}=\pm\frac{3\sqrt{5}}{2} H\left[\hat{\mathcal{F}}_{B-} ^{\dagger}\hat{\mathcal{F}}_{B-}+\hat{\mathcal{F}}_{A-} ^{\dagger}\hat{\mathcal{F}}_{A-} \right].  \label{37}
\end{eqnarray} 
Notably the vacuum state $\ket{0}$ is same for both A and B type particle in both the systems.
\par  The simultaneous particle creation - annihilation mechanism in the ( Universe + Anti-Universe) and corresponding Parallel system has been demonstrated in a diagram in Fig-$1$ and Fig-$2$. In an expanding Universe ($H>0$), a A-particle with energy $+\frac{3\sqrt{5}}{2}H$  and momentum $+p$ is annihilated and simultaneously an anti A-Particle with energy  $-\frac{3\sqrt{5}}{2}H$ and momentum $-p$ is created in the contracting PU. Again in the expanding Universe a B-particle with energy  $+\frac{3\sqrt{5}}{2}H$ and momentum $+p$ is created when an anti B-particle with energy  $-\frac{3\sqrt{5}}{2}H$ and momentum $-p$ is annihilated in the contracting PU. Similarly in an expanding anti Universe, an anti A-particle with energy $+\frac{3\sqrt{5}}{2}H$ and momentum $-p$ is annihilated and a anti B-particle with energy $+\frac{3\sqrt{5}}{2}H$, momentum $-p$ is created while in the contracting parallel of the Anti Universe,  a B-particle with energy $-\frac{3\sqrt{5}}{2}H$, momentum $+p$ is annihilated and a A-particle with energy $-\frac{3\sqrt{5}}{2}H$, momentum $+p$ is created.

\par In this work, we assumed the system of real scalar field to describe the dynamics of the combination of the Universe- Anti  Universe and its Parallel counter part. The A and B type particles are charge less scalar Bosons.  Clearly, the creation and annihilation of respective B-particle and A-particle in an expanding Universe are entangled by the annihilation and creation of A and B type anti particles with opposite momenta in the expanding AU. This whole process has a CPT- invariant contracting Parallel system(which includes the Parallel Universe and the Parallel of the AU). This whole evolution picture has been depicted in the Fig.$1$ and Fig.$2$ qualitatively. 

\subsection{Symmetry Breaking near the emergent phase $H\rightarrow0$ and mass generation in the scalar Bosons.}
The general form of the scalar field at any epoch is mentioned in the equation (\ref{14}). In order to normalize the scalar field, we take 
\begin{equation*}
		\hat{\phi}(X)=\frac{1}{2}\left [\hat{\phi}_{+}(X)+\hat{\phi}_{-}(X) \right]	.
\end{equation*}  
At the emergent epoch($a\rightarrow a_E, H=0$), one finds $\hat{\phi}_{+}(X)\xrightarrow[]{ H=0} a_E^{\frac{3}{2}}\phi _{0+}, ~ \hat{\phi}_{-}(X)\xrightarrow[]{ H=0} a_E^{-\frac{3}{2}}\phi _{0-} $. $\phi _{0+}=\phi _{0-}=\phi_0$ is the free real scalar field. For simplicity of calculation, we choose in a suitable unit system ,$ a_E =1$. Hence one can find 
$\hat{\phi}(X)\xrightarrow[]{ H=0} \phi _{0}$. Clearly the system behaves as a free scalar field system at the emergent epoch with the Lagrangian
\begin{equation*}
	\mathcal{L}_0= \frac{1}{2}\left [\dot{\phi}_o^2-|\vec{\nabla} \phi_{0}|^2-m_0^2 \phi_{0}^2 \right ],
\end{equation*}
where $m_0$ is mass of the scalar bosons at the emergent epoch. At any arbitrary epoch, $\phi _{0\pm}\rightarrow \phi_{\pm}=a^{\pm \frac{3}{2}}\phi_0$. Considering the epoch nearing the emergent era($a$ is sufficiently small), one has $\phi_{\pm}\simeq\phi_{0}(1+\pm a^{\frac{3}{2}}+\frac{1}{2} a^3)$. Hence effectively we obtains the form of the scalar field at the very early epoch of evolution as 
\begin{equation*}
	\phi=\phi_{0}\left (1+\frac{1}{2}a^3\right ).
\end{equation*} 
Notably we take up to the second order term of the exponential series to contain the least non-trivial change in the field. There is no change in the total scalar field due to the first order term. Consequently the expression of the Lagrangian takes the form
\begin{equation*}
	\mathcal{L}=\frac{\left (1+\frac{1}{2}a^3 \right )^2}{2}\left [\dot{\phi}_o^2-|\vec{\nabla}^{\prime} \phi_{0}|^2-\left\{ m^2 -\left(\frac{\frac{3}{2}a^3 H}{1+\frac{1}{2}a^3}\right)^2\right \} \phi_{0}^2  +2\left(\frac{\frac{3}{2}a^3 H}{1+\frac{1}{2}a^3}\right) \phi_{0} \dot{\phi_{0}}\right ].
\end{equation*}  
 This form of the Lagrangian contains the effective mass term  $ m^2 -\left(\frac{\frac{3}{2}a^3 H}{1+\frac{1}{2}a^3}\right)^2$.
If we look back to the expression of momenta of the A and B particles, it approaches to zero at the origin epoch with mass $m_0=0, \omega_0=0$. Hence there is no contribution of the A and B particles to the total energy at the emergent phase. Hence one can ignore the relevance of these scalar bosons at the emergent epoch. This phase is completely dominated by the  free real scalar field particles.  At the termination  of the emergent phase, there exists a spontaneous Lorentz symmetry breaking which causes the generation of this two types of scalar field particles. One can investigate the variation of mass of the particles by studying this symmetry breaking mechanism..

 Here we already set the model with constant comoving momenta $p$. So we obtain the variable mass of the scalar field particle as 
\begin{equation*}
	m(H)=\frac{9}{4}H^2 +\left(\frac{\frac{3}{2}a^3 H}{1+\frac{1}{2}a^3}\right)^2-p^2 .
\end{equation*}
Thus this spontaneous symmetry breaking process near the emergent origin may be the possible mechanism behind the accelerated expansion just after the emergent scenario.

	\begin{figure}[h]
	\begin{minipage}{0.9\textwidth}
		\centering
		\includegraphics*[width=0.9\linewidth]{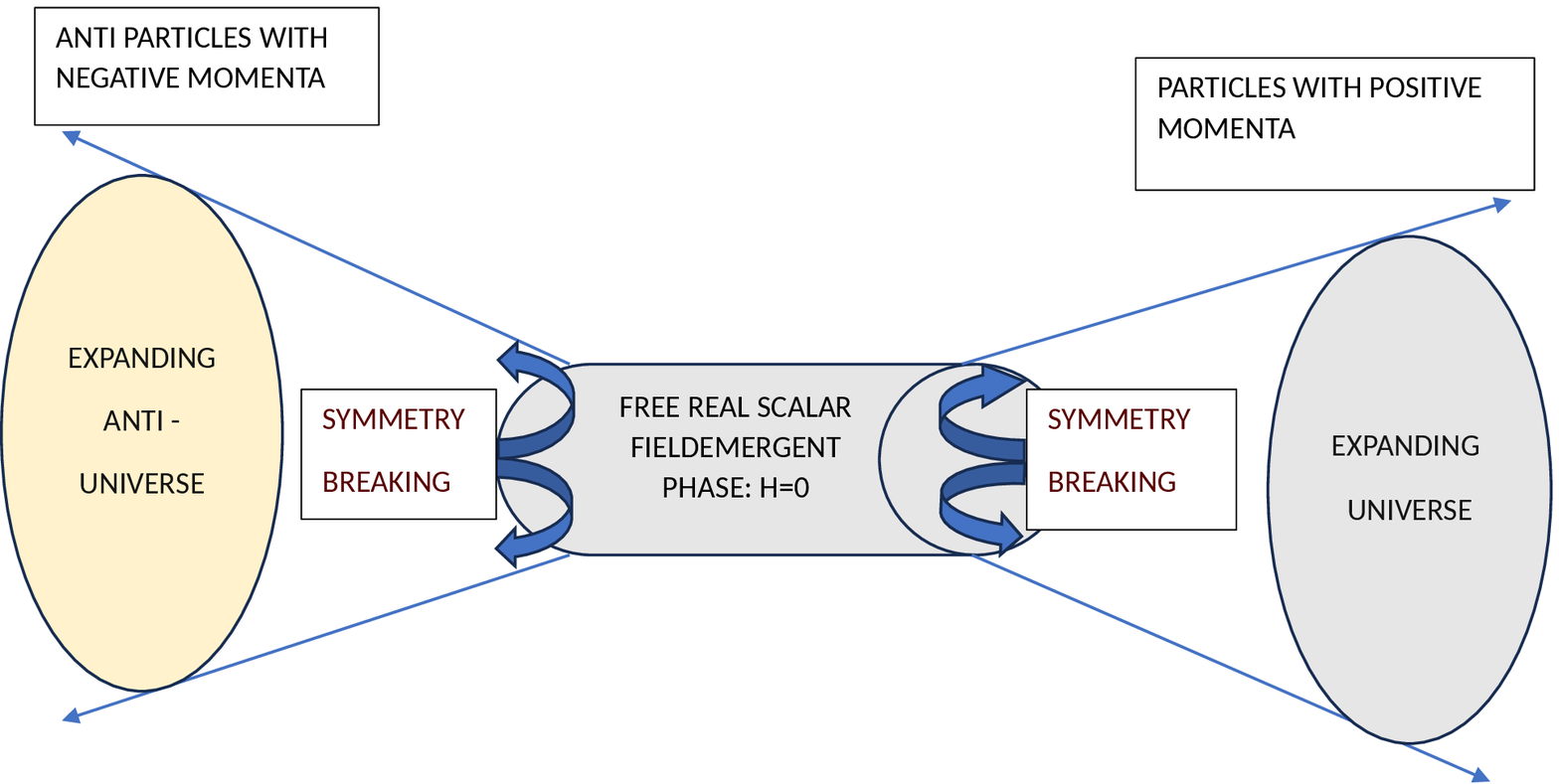}\\
		(a)
	\end{minipage}
	\begin{minipage}{0.9\textwidth}
		\centering
		\includegraphics*[width=0.9\linewidth]{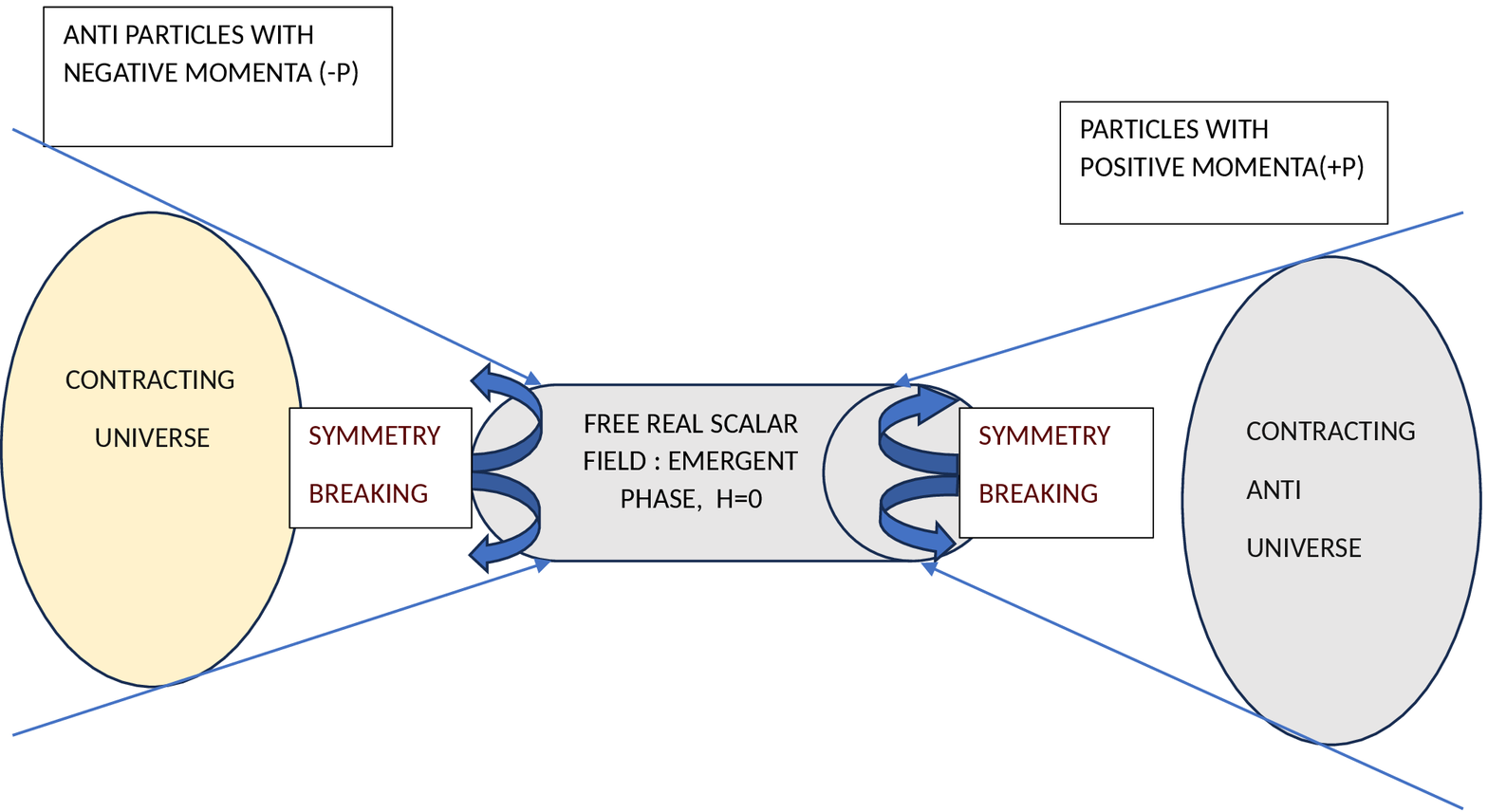}\\
		(b)
		
	\end{minipage}
	\begin{center}
		\caption{ (a)Expanding Universe and expanding Anti-Universe       (b) Parallel system of the expanding Universe$+$ Anti-Universe system.} 
	\end{center}
	\label{fig2}
\end{figure}

\section{Expansion  of the Universe  and the Anti Universe under quantized cosmic fluid }

As per the outcome of the previous section, we have the creation and annihilation mechanism of two types of particle simultaneously. Type - A particles are annihilated and type - B particles are created with evolution of the Universe. The energy of each of both types of particles are same and varies linearly with the Hubble parameter. The number operators are 
$\hat{N}_A= \hat{\mathcal{F}}^{\dagger}_A\hat{\mathcal{F}}_A, \hat{N}_B = \hat{\mathcal{F}}^{\dagger}_B\hat{\mathcal{F}}_B$

 In an isolated Universe, total energy is conserved and hence, one has
\begin{equation}
	(N_A+N_B)\omega_0= \mbox{Constant,}  \label{38}
\end{equation} $N_A, N_B$ are the expectation values of $\hat{N}_A$ and $\hat{N}_B$ respectively.
Again
$\dot{N}_A=-3H N_A, \dot{N}_B=3H N_B$ . There fore equation (\ref{38}) leads to
\begin{equation}
	(\dot{H}-3H^2) N_A +(\dot{H}+3H^2)N_B=0 . \label{39}
\end{equation}
The equation (\ref{39}) can be written in the form
\begin{equation}
	\dot{H}+3\frac{N_B- N_A}{N}H^2=0 , \label{40}
\end{equation}
$N=N_A+N_B$, the total number of particles at an epoch when the Hubble parameter is $H$.

Then one can write,
\begin{equation}
	\dot{H}-3H^2\left (P_A(H)-P_B(H)\right )=0  \label{41} 
\end{equation} where $P_A(H), P_B(H)$ are the occupancy probabilities of the two types of particles respectively at any particular Hubble parameter value $H$.  Also $P_A(H)+P_B(H)=1$. This is the form of the governing equation of the evolution of the Universe. The phenomenological choice of the occupancy probability can be used to describe the cosmic evolution pattern following this evolution equation.

\section{Different phases of cosmic evolution : a phenomenological approach.}

In this section, it is aimed to find whether there is a  choice of the continuous occupancy probability function which can satisfy the complete cosmic evolution from the emergent era to present late time acceleration phase.
Evidently, the dependence of the occupancy probability of one of the two types of particle on the Hubble parameter will be of different natures in different cosmic phases like emergent, inflation, decelerating expansion and late time acceleration. For continuous evolution, all the parameters including the occupancy probability will be continuous across the transition epochs  of the consecutive eras.
\par The evolution equation (\ref{41}) can be written as
\begin{equation}
	\dot{H}+3H^2\left [1-2P_A(H)\right ]=0.  \label{42}
\end{equation}
 Here  the Universe is statistically an instantaneous micro-canonical ensamble with the parameters $(E=\omega_0= \frac{3}{2}H, N= N_A+N_B, V=V_0 a^3)$. All the macrostate parameters are evolving with time. But considering the instantaneous equilibrium at an epoch $(t)$, The occupancy probability within an energy range centered at $E=\frac{3}{2}H$ is given by $P_A(H)=\frac{1}{W_A(H)}$, where $W_A(H)$ is the number of micro-states under that particular macrostate. 

Hence we can start with a formal choice of the form  of the occupancy probability as a power law of energy, 
\begin{equation}
	P_A(E)\propto E^n \implies P_A(H)\propto H^n .   \label{43}
\end{equation}

Then the values of $n$ can be chosen phenomenologically to match with the desired pattern of the cosmic evolution.
\par $(i)$ $P_A(H)=\frac{e}{2 H}$ for emergent phase in the range of epoch $t\leq t_0$.
\par $(ii)$ $P_A(H)=\frac{I}{2}H$ for inflation in the range $t_0\leq t \leq t_1$.
\par $(iii)$ $P_A(H)=\frac{d}{2}$ (\mbox{Constant}) for decelerated expansion in the range $t_1\leq t \leq t_3$.
\par $(iv)$ $P_A(H)=\frac{L}{2 H^2}$ for late time acceleration phase in the range $t\geq t_3$.  ~~$e,I,d,L$ are pure constants and  their dimensions are suitably  adjusted.
 
\subsection{Emergent scenario :}
 Here the equation(\ref{42}) is found in the form,
 \begin{equation}
 	\dot{H}+3H(H-e)=0 .  \label{44}
 \end{equation}
The solutions are in the form
\begin{equation}
H=\frac{e}{1+(\frac{e}{H_0}-1)\exp[-3e(t-t_0)]}.	 \label{45}
\end{equation}			
\begin{equation}
	a=\left[\frac{\left (\frac{e}{H_0}-1 \right )+\exp\left(3 e (t-t_0)\right)}{\frac{e}{H_0}}\right]^{\frac{1}{3}} \implies H=e+\frac{H_0-e}{a^3} . \label{46}
\end{equation}	
In the condition $e>H_0$, there is no singularity in the real time and also it satisfies the criteria of emergent scenario :
\par 
$H\rightarrow 0, a\rightarrow\left[\frac{e-H_0}{e}\right]^{
\frac{1}{3}
}$, when $t\rightarrow -\infty$.
\par 
$H\rightarrow 0, a\rightarrow\left[\frac{e-H_0}{e}\right]^{
	\frac{1}{3}
}$, when $t<< t_0$.
\par 
$H\sim e, a\simeq \frac{H_0}{e}\exp(3e(t-t_0))$, when $t>>t_0$.
Here $t_0$ is a reference epoch of time. Here we choose it as the starting point of the inflationary era.
\subsection{Inflationary era :}
In this case, the evolution equation (\ref{42}) takes the form
\begin{equation}
	\dot{H}+3H^2(1-IH)=0.   \label{47}
\end{equation}
The suitable solution are 
\begin{equation}
	H=\frac{H_1}{IH_1+(1-IH_1)(\frac{a}{a_1})^3}  \label{48}
\end{equation} and
\begin{equation}
	a=a_1\left[\frac{IH_1}{1-IH_1}\mbox{Lambert}W \left(\frac{1-IH_1}{IH_1}\exp \frac{(1-IH_1)+H_1(t-t_1)}{IH_1}\right)\right]^{\frac{1}{3}}.  \label{49}
\end{equation}
Here $a_1=a(t_1),H_1=H(a_1)$ and for smooth evolution, $IH_1<1$ must be satisfied.
The deceleration parameter is found as
$q=-IH$. Hence $I$ must be positive quantity for accelerated expansion. So the restriction on the parameter $I$ in this model is $0<I<\frac{1}{H_1}$. The solution also satisfies the criterion of early time exponential acceleration $a<<a_1, H\rightarrow \frac{1}{I}$, a constant.
\subsection{Decelerating expansion :}
In this cosmic phase, the explicit form of evolution equation (\ref{42}) is
\begin{equation}
	\dot{H}+3H^2(1-d)=0  . \label{50}
\end{equation}
The solutions are found 
\begin{equation}
	H=H_1\left[\frac{a}{a_1}\right]^{3(d-1)}.\label{51}
\end{equation}
\begin{equation}
	a=a_1\left[1+3H_1(1-d)(t-t_1)\right ]^{\frac{1}{3(1-d)}}.  \label{52}
\end{equation}
Here the deceleration parameter is found as
$q=(2-3d)$. So for deceleration, $d<\frac{2}{3}$ is the restriction on the choice of $d$.
\subsection{Late time acceleration :}
Here the Evolution equation (\ref{42}) yields
\begin{equation}
	\dot{H}+3(H^2-L)=0 .\label{53}
\end{equation}

The solutions are 
\begin{equation}
	H= \sqrt{L+(H_3 ^2-L)\left(\frac{a}{a_3}\right)^{-6}}. \label{54}
\end{equation}
\begin{equation}
	a=a_3\left[\sqrt{\frac{H_3^2-L}{L}}\sinh\{3\sqrt{L}(t-t_j)\}\right]^{\frac{1}{3}}.
		\label{55}
\end{equation}

Here $a_3=a(t_3), H_3=H(a_3)$. $t_j$ is a reference of epoch of time. Here one has $t_3-t_1 =\frac{1}{3(1-d)}\left(\frac{1}{H_3}-\frac{1}{H_1}\right)$ and $3\sqrt{L}(t_3-t_j)=\sinh^{-1}\sqrt{\frac{d}{1-d}}$ . The deceleration parameter in this case is $q=3\left(1-\frac{L}{H^2}\right)-1$. Hence for accelerating expansion, the condition on the choice of $L$ is $\frac{2}{3}H_3^2<L<H_3^2$.

\par The solutions of evolution equation for different phases has been represented in the Table :$1$ with the suitable restrictions on the parameters. Also a complete and continuous cosmic evolution pattern with continuous variation of different parameters has been presented graphically in Fig.$3$.

 \begin{center}
	\begin{table}[!htb]
		\centering
		\renewcommand{\arraystretch}{2.5}
		\caption{FORM OF $P_A(H)$ AND RESTRICTION ON CONSTANTS AT DIFFERENT ERA OF EVOLUTION .} \label{tab:1}
		~~~~~~~\begin{tabular}{| >{\centering\arraybackslash}m{2.5cm}|>{\centering\arraybackslash}m{10cm}|>{\centering\arraybackslash}m{5cm}|}
			\hline
			Form of $P_A(H)$ &  Explicit solution of the parameters & restricted range of constants for valid evolution \\
			\hline
			$P ^{(e)}_A(H) \propto \frac{1}{H}$ & $H^{(e)}=\frac{e H_0}{H_0+(e -H_0)\exp{-3e(t-t_0)}} $  & $e>H_0$ for emergent scenario. \\
			&$a^{(e)}=\left[\frac{\left (\frac{e}{H_0}-1 \right )+\exp\left(3 e (t-t_0)\right)}{\frac{e}{H_0}}\right]^{\frac{1}{3}}$&  \\
			
			\hline
			$P^{(I)}_A(H) \propto H$ & $H^{(I)}=\frac{H_1}{IH_1+(1-IH_1)(\frac{a}{a_1})^3} $ & $ 0<I<\frac{1}{H_1} $  for early time acceleration.\\ 
			&$a^{(I)}=a_1\left[\frac{IH_1}{1-IH_1}\mbox{Lambert}W \left(\frac{1-IH_1}{IH_1}\exp \frac{(1-IH_1)+H_1(t-t_1)}{IH_1}\right)\right]^{\frac{1}{3}}$&\\
			&$q^{(I)}=-IH^{(I)}$&\\
			\hline
			$P^{(D)}_A(H) \propto H^0$& $H^{(D)}=H_1 \left(\frac{a}{a_1}\right)^{3(d -1)}$ & $d<\frac{2}{3}$  for decelerating expansion. \\ 
			&$a^{(D)}=a_1\left[1+3H_1(1-d)(t-t_1)\right ]^{\frac{1}{3(1-d)}}$&\\
			&$q^{(D)}=(2-3d)$& \\
			\hline 
				$P^{(L)}_A(H) \propto H^{-2}$ & $H^{(L)}=\sqrt{L+({H_3}^2-L)\left(\frac{a}{a_3}\right)^{-6}} $ & $\frac{2}{3}{H_3}^2<L<{H_3}^2 $ for late time acceleration. \\
				&$a^{(L)}=a_3\left[\sqrt{\frac{H_3^2-L}{L}}\sinh\{3\sqrt{L}(t-t_j)\}\right]^{\frac{1}{3}}$&\\
				&$q^{(L)}=3\left(1-\frac{L}{H^2}\right)-1$&\\ 
			\hline
		\end{tabular}
	\end{table}
\end{center} 
\par From the continuity of the occupancy probability $P_A(H)$ across the two transition epochs $t_1$  and $t_3$ yield 
\begin{equation}
	d=IH_1.  \label{56}
\end{equation}
\begin{equation}
	L=dH_3^2  . \label{56}
\end{equation}

Also we have $H_3=H_1\left(\frac{a_3}{a_1}\right)^{3(d-1)}$.

\begin{figure}[h]
	\begin{minipage}{0.49\textwidth}
		\centering
		\includegraphics*[width=0.9\linewidth]{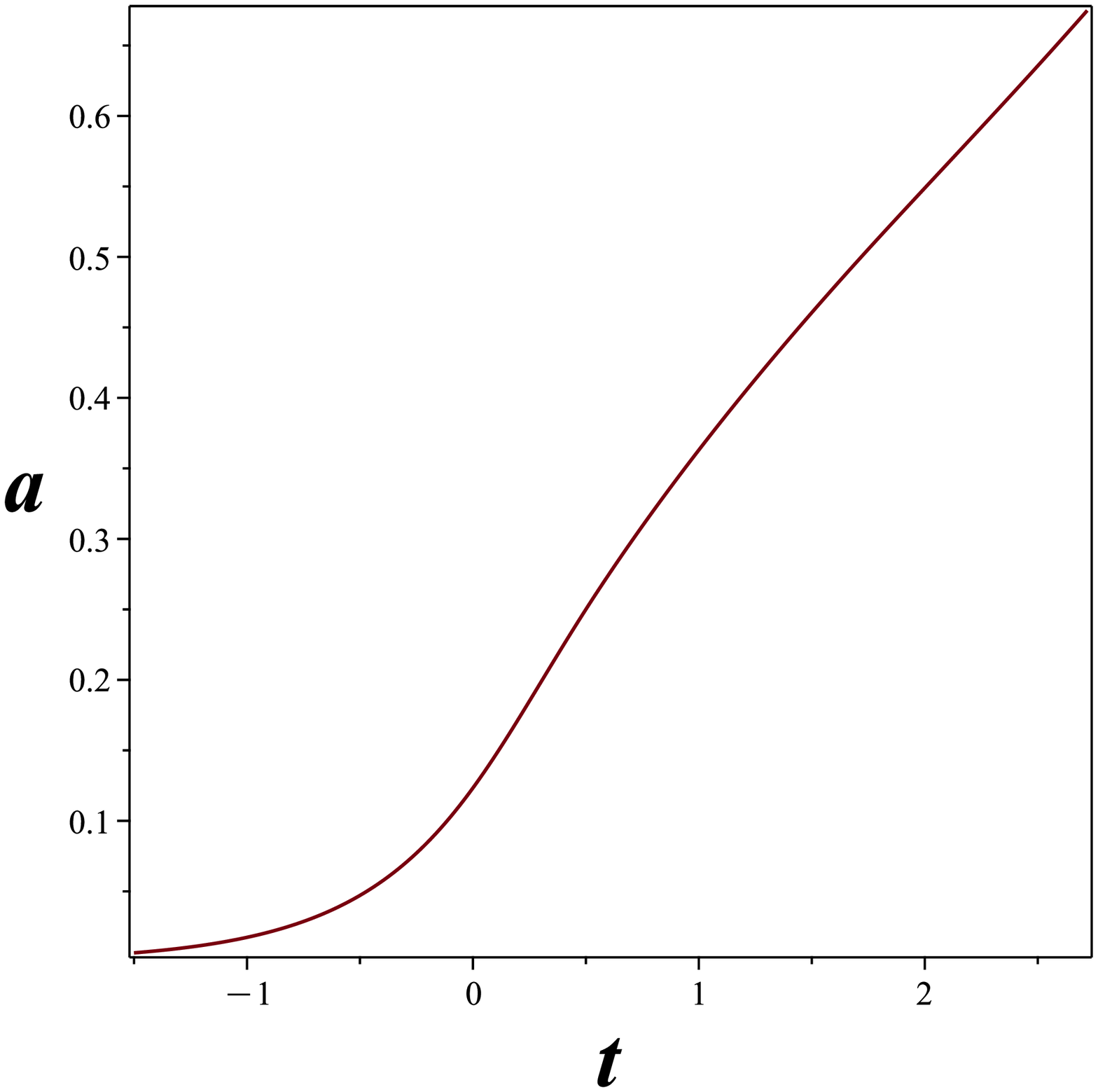}\\
		(a)
	\end{minipage}
	\begin{minipage}{0.49\textwidth}
		\centering
		\includegraphics*[width=0.9\linewidth]{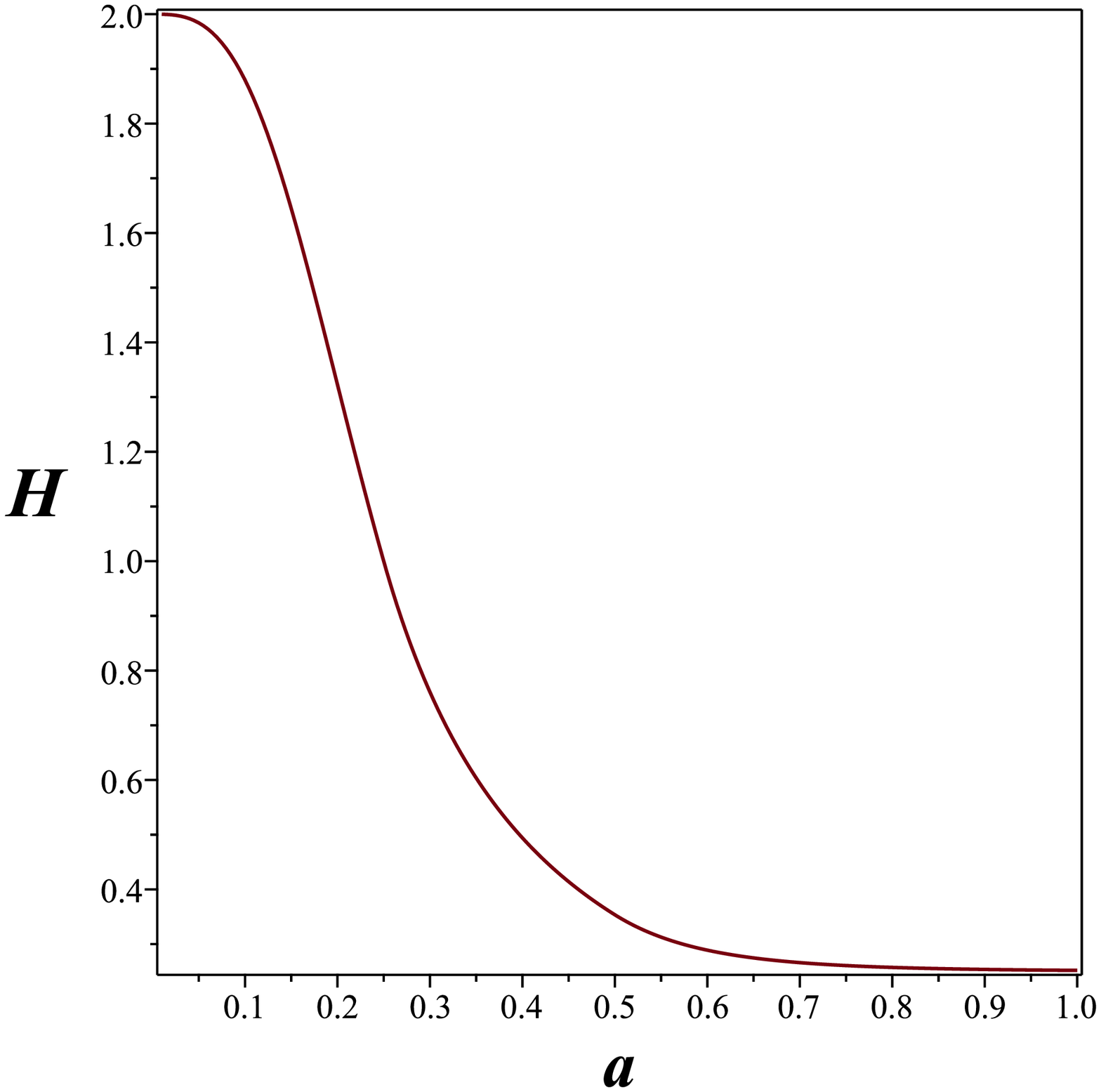}\\
		(b)
		
	\end{minipage}
\begin{minipage}{0.49\textwidth}
	\centering
	\includegraphics*[width=0.9\linewidth]{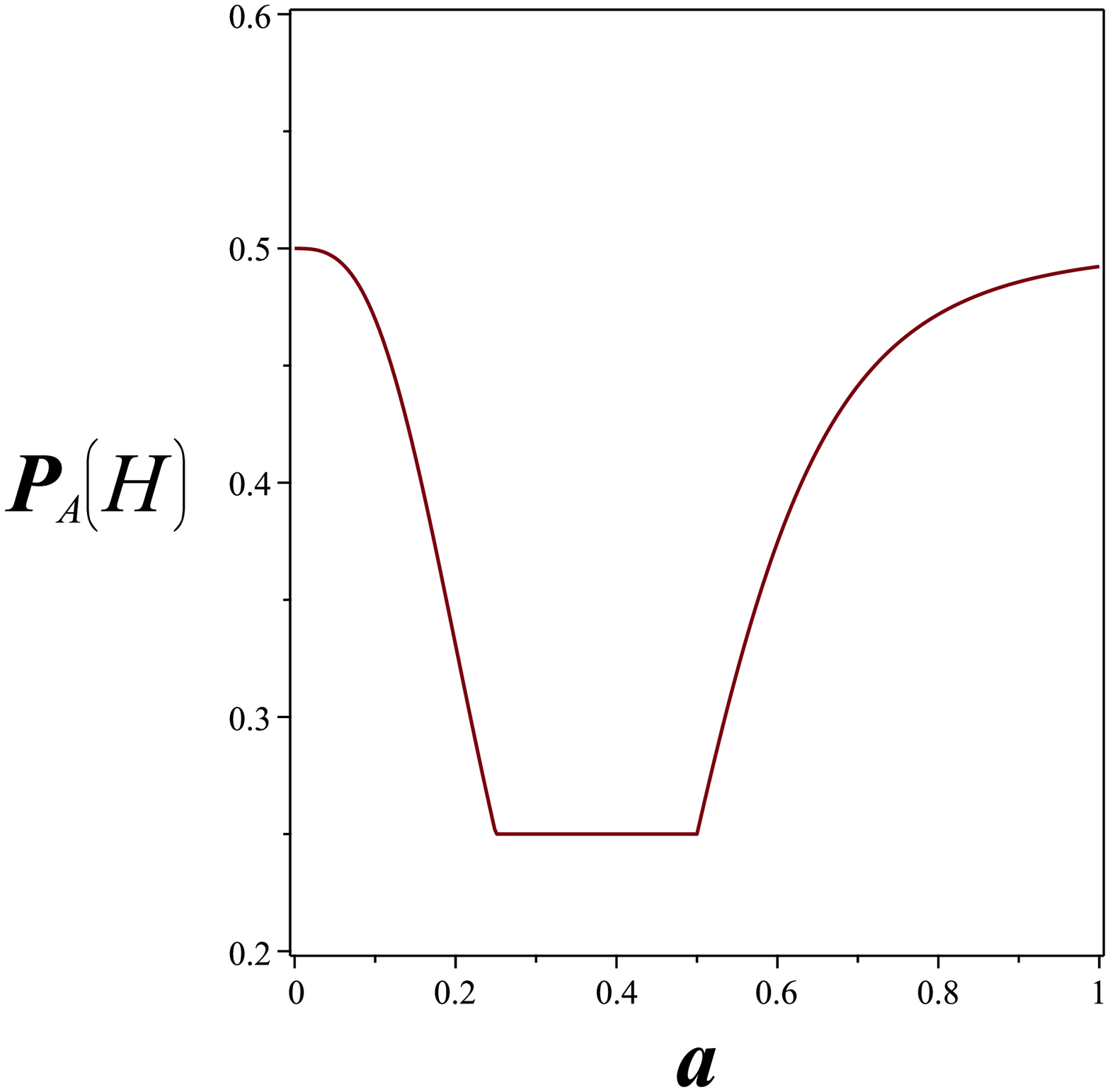}\\
	(c)
	
\end{minipage}
\begin{minipage}{0.49\textwidth}
	\centering
	\includegraphics*[width=0.9\linewidth]{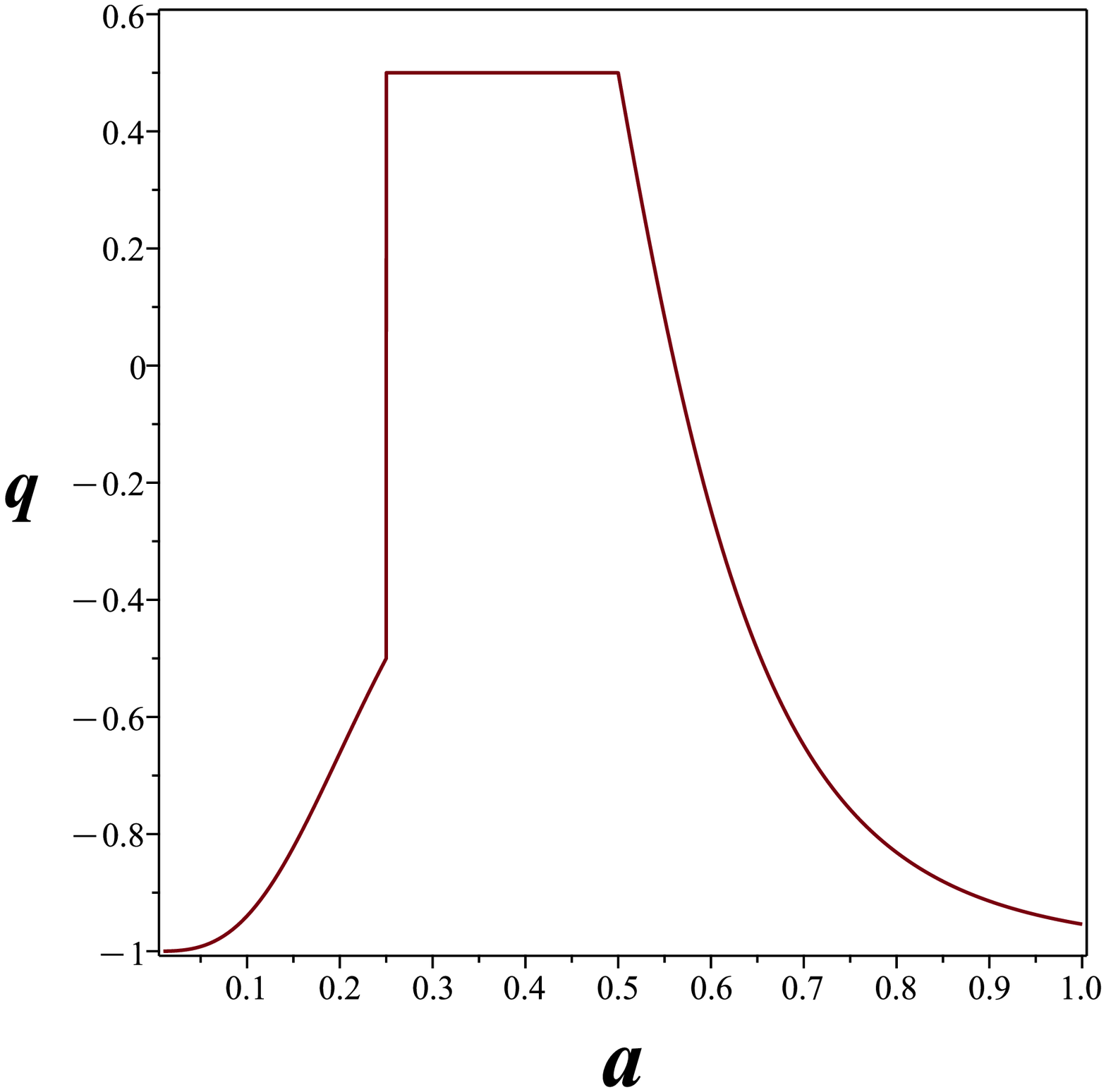}\\
	(d)
	
\end{minipage}
	\begin{center}
		\caption{ (a)Variation of Scale factor  : $a$ with time $t$   (top left)   (b) Variation of Hubble parameter : $H$with $a$ (top right), (c) $P_A$ with scale factor $a$ (bottom left) and (d)deceleration parameter $q$ with $a$(bottom right)  for  $a_1=0.25,t_1=0.5,H_1=1,a_3=0.5,I=0.5$ .} 
	\end{center}
	\label{fig3}
	
\end{figure}

\section{Discussion}

The approach of this work is to find the cosmic evolution equation from the quantum dynamics of the cosmic fluid. Here the cosmic fluid is chosen as the real scalar field Lagrangian with some suitable periodic source term. The choice of the source term is so adjusted to get booth the damped and growing oscillator components of the real scalar field. Hence the process of such canonical quantization describes the stimulated particle creation-annihilation mechanism. But it is important to mention that the extra source term ( in the KG equation) can not be provided externally in an adiabatic Universe. This source arises as a consequence of the entanglement between the evolution of the Universe and its Anti-Universe. This modified form of the real scalar field describes the cosmic fluid of the "Universe + Anti-Universe" system. The whole system is CPT-invariant i.e. it corresponds to the existence of a parallel system.

  Following this prescription, the solution is found to have two types of particles with their consecutive anti-particles. As the scalar bosons are charge less and spin $0$ particles, the respective antiparticles will also be charge less and  spin $0$. The particles will occupy the Universe and the parallel of the Anti-Universe. The anti particles will occupy the Anti- Universe and the parallel Universe.   The entanglement between the Universe and its Anti-Universe leads to the simultaneous pair creation-annihilation  with opposite momenta.
  \par This model is found to be a non-singular model of the Universe. At the origin, there exists a free real scalar field particles which yields an  emergent cosmic phase. At the termination of this state, a spontaneous Lorentz symmetry breaking process occurs which stimulates the system to generate the damped and growing scalar bosons. The symmetry breaking phenomena yield the bosons with time varying mass.  
  
   \subsection{Equivalence of the quantization evolution and GTR : an alternative to the dark energy. } However the evolution pattern of the Universe is found to be dependent on the occupancy probability of the cosmic fluid particles.
\par The evolution equation equation (\ref{41}) can be written in the from of the evolution equation in Friedmann cosmology $\dot{H}+\frac{3}{2}H^2(1+\omega_f)=0$ where $\omega_f$ is the effective barotropic index of the equivalent barotropic cosmic fluid in Friedmann Universe. Here $\omega _f=1-4P_A(H)$. Evidently, the equivalent cosmic fluid will have dark energy component when its occupancy probability will be greater than $\frac{1}{4}$.  Hence one may conclude that under an instantaneous Micro-canonical ensamble, the mixture of two types of cosmic fluid particles (one (B) under creation process and another (A) under annihilation ) can act as a dark fluid when  $P_A\geq \frac{1}{4}, ~ P_B \leq \frac{3}{4}$.

\par Finally, following the suitable choices for the occupancy probability of the A-type particle, we have successfully presented a continuous and complete evolution from non-singular emergent phase to present late time acceleration era through two intermediate eras inflation and decelerating expansion.

\par In this model, the late time acceleration phase will  exist for ever. It will prevail  until the occupancy probability $P_A$ reaches its maximum value $1$. But this condition never be achieved in any real time.  Hence the duration of the late time acceleration phase has no limit unless it changes the evolutionary pattern.

\par Notably, the emergent era of this model (the Universe consists of two types of particles-A,B ) is completely non-singular i.e. with out any big-rip or big bang singularity. This result is unlike our previous work\cite{Maity:2023pmx}, where there was a mathematical big-rip singularity beyond the valid epoch of time in a Universe with one type of particle . The equation(\ref{20})  can be written as 
$\frac{H^2}{H^2_0}=\Omega _{\Lambda _0}+\Omega_M(1+z)^{3(1+\omega_{M})}+\Omega_{\mbox{MP}}(1+z)^{3\left (1+\omega _{\mbox{MP}}\right )} $,where $z=\frac{1}{a}-1$, the cosmological red shift, $\Omega _{\Lambda _0}=\frac{e^2}{H^2_0},\Omega_M=2e(H_0-e)\frac{1}{H^2_0} , \Omega_{\mbox{MP}}=\left(1-\frac{e}{H_0} \right)^2$.Here $\Omega_{\Lambda _0}+\Omega_M+\Omega_{\mbox{MP}}=1$. Effectively $\omega_M =0, \omega _{\mbox{MP}}=1$. Clearly at, $z\rightarrow -1,a\rightarrow \infty$, the present model 
leads to the de sitter expansion which $\Lambda$- CDM also has in this epoch.

\par  This work provides a qualitative mechanism behind the origin and the complete - continuous evolution of the Universe. The authors admit that further studies may yield the better fine tuning of the parameters used in this model. Especially we have ignored the gauge interaction of Bosons with the cosmic fluid and hopefully in future works, we shall reflect on the  further modification of this model.  

Besides, it can  be examined whether there is a correspondence  between the thermodynamics of the cosmic fluid and its quantum dynamics in the cosmological perspective.

\section*{Acknowledgment } The author SM thanks to Prof. Subenoy Chakraborty, Dept. of Mathematics, Jadavpur University, Kolkata-$700032$ for his valuable suggestion on  this topic and the author MD acknowledges University Grant Commission, Govt. of India for providing Junior Research fellowship during this work.

       \end{document}